\documentclass[aps,prb,showpacs,twocolumn]{revtex4}
\usepackage{bm,color,amsmath,amssymb,mathrsfs,latexsym,graphicx,psfrag}

\newcommand{\bs}[1]{\boldsymbol{#1}}

\renewcommand{\i}{\text{i}}

\begin{document}
\title{Generalizations of Perelomov's identity on the completeness of
    coherent states}
\author{Martin Greiter}
\affiliation{Institut f\"ur Festsk\"orperphysik, Karlsruhe Institute of
  Technology, 76021 Karlsruhe, Germany}
\author{Ronny Thomale}
\affiliation{Department of Physics, Stanford University, Stanford, CA 94305, 
  USA}
\pagestyle{plain}

\date{\today}

\begin{abstract}
  We proof the Perelomov identity for arbitrary 2D lattices using Fourier
  transformation.  We further generalize it to situations where the origin
  does not coincide with a lattice site, and where the form of the exponential
  factor is reminiscent of magnetic wave functions in uniaxial rather than
  symmetric gauge.
\end{abstract}

\pacs{75.10.Jm,02.30.Lt,02.30.Fn}

\maketitle
\emph{Introduction.}---Studies of two-dimensional spin
liquids~\cite{anderson87s1196,kalmeyer-87prl2095,kivelson-87prb8865,wen-89prb11413,zou-89prb11424,kalmeyer-89prb11879,moessner-01prl1881,balents-02prb224412,greiter02jltp1029,kitaev06ap2,schroeter-07prl097202,yao-07prl247203,dusuel-08prb125102,lee08s1306,greiter-09prl207203,thomale-09prb104406,hermele-09prl135301,balents10n199},
which were originally motivated by the problem of high $T_{\rm c}$
superconductivity, are enjoying a renaissance of interest in present
days~\cite{zhang-11prb075128,grover-11prl077203,yao-11prl087205,scharfenberger-11prb140404,nielsen-1201.3096,greiter-1201.5312}.
One of the reasons is the accumulation of numerical evidence for spin
liquid phases in various two dimensional spin models, including the
Hubbard model on a honeycomb lattice~\cite{meng-10n847} and the
next-nearest neighbor Heisenberg antiferromagnet on the square
lattice, usually referred to as the $J_1$--$J_2$
model~\cite{jiang-11arXiv:1112.2241}.  Another reason is that they
constitute intricate examples of topological
phases~\cite{wen89prb7387,wen90ijmp239,wen-90prb9377,wen04,levin-05prb045110,isakov-11np772},
which currently receive significant interest in the context of
topological
insulators~\cite{moore10n194,hasan-10rmp3045,qi-11rmp1057}.

In fact, the concept of topological
order~\cite{wen89prb7387,wen90ijmp239,wen-90prb9377,wen04} was
discovered in a two dimensional spin liquid, the (Abelian) chiral spin
liquid
(CSL)~\cite{kalmeyer-87prl2095,wen-89prb11413,zou-89prb11424,kalmeyer-89prb11879,laughlin-90prb664,schroeter-07prl097202,thomale-09prb104406,nielsen-1201.3096,greiter-1201.5312}.
The idea of this spin $s=\frac{1}{2}$ liquid, due to
D.H.~Lee~\cite{laughlinPC}, is to describe spin flip operators $S^+_i$
in a background of down spins by a bosonic quantum Hall wave function
at Landau level filling factor $\nu=\frac{1}{2}$.  Kalmeyer and
Laughlin~\cite{kalmeyer-89prb11879} discovered that this wave
function, when supplemented by an appropriate gauge factor $G(z)=\pm
1$, is a spin singlet.  As first pointed out by Zou, Dou\c cot, and
Shastry~\cite{zou-89prb11424}, the proof relies on an identity
established by A.M.~Perelomov in 1971 in the context of the
completeness of systems of coherent states~\cite{perelomov71tmp156}.
The Abelian CSL is the simplest example of a class of spin liquids,
which are constructed using Landau level wave functions in ficticious
or auxiliary magnetic fields.  More recent examples of this class
include the spin $S=1$ chirality liquid~\cite{greiter02jltp1029}
(which is constructed via Schwinger boson projection form two Abelian
CSLs with opposite chirality), the spin $s$ non-Abelian
CSL~\cite{greiter-09prl207203,greiter-1201.5312} (which supports
spinon excitations with SU(2) level $k=2s$ statistics), and a
hierarchy of spin liquid states~\cite{scharfenberger-11prb140404}
(which suggests that spinons in parity and time reversal invariant
antiferromagnets with integer spin $s=2$ and higher obey SU(2) level
$k=s$ non-Abelian statistics).

All the spin liquids in this class share two features.  First, the
mechanism of fractional quantization yielding spinon (and holon)
excitations, is both mathematically and conceptually similar to the
mechanism of fractional quantization in quantized Hall states.  The
fractional quantum number in the spin liquids is the spin
$s=\frac{1}{2}$ of the spinon, which is fractional in the context of
Hilbert spaces built out of spin flips, which carry spin one.  The
Abelian CSL is related to a Laughlin state in the quantum Hall system,
while the family of non-Abelian CSLs are reminiscent of the
Moore--Read~\cite{moore-91npb362,greiter-92npb567} and Read--Rezayi
states~\cite{read-99prb8084}.  Second, many analytical results
available for these highly complex states, including the singlet
property for Abelian~\cite{kalmeyer-89prb11879,zou-89prb11424} and
non-Abelian CSL states~\cite{greiter-09prl207203}, as well as the
recent construction of a parent Hamiltonian~\cite{greiter-1201.5312}
for the non-Abelian CSL states, rely on Perelomov's
identity~\cite{perelomov71tmp156}.

This identity was originally derived from the properties of the Jacobi
theta functions, and used to show that there is only one linear
relation between certain systems of coherent states.  In this Brief
Report, we show that Perelomov's identity can be generalized or
reformulated in several ways, which are highly expedient for applications
to spin liquids.

To be precise, we do three things.  First, we proof that the identity
holds for arbitrary 2D lattices with one site per unit cell using
Fourier transformation.  Second, we generalize the identity to
situations where the origin does not coincide with a lattice site.
Third, we rewrite the identity such that the form of the exponential
factor is reminiscent of magnetic wave functions in uniaxial rather
than symmetric gauge.  The last result is particularly useful when the
spin liquids are formulated on lattices with periodic boundary
conditions.


\vspace{10pt} 
\emph{The Perelomov identity}.---Consider a
lattice 
spanned by $\eta_{n,m}=na+mb$ in the complex plane, with $n$ and $m$ integer
and the area of the unit cell $\Omega$ spanned by the primitive lattice
vectors $a$ and $b$ set to $2\pi$,
\begin{equation}
  \label{eq:area}
  \Omega=\left|\Im(a\bar b)\right|
  = 2\pi,
\end{equation}
where $\Im$ denotes the imaginary part.  Let $G(\eta_{n,m})=
(-1)^{(n+1)(m+1)}$.  Then
\begin{equation}
  \label{eq:perel}
  \sum_{n,m} P(\eta_{n,m}) G(\eta_{n,m}) e^{-\frac{1}{4}|\eta_{n,m}|^2}=0
\end{equation}
for any polynomial $P$ of $\eta_{n,m}$.

\vspace{3pt}
\emph{Proof.}---It is sufficient to proof the identity for the
generating functional
\begin{equation}
  \label{eq:genfun}
  \sum_{n,m} e^{\frac{1}{2}\eta_{n,m}\bar z} G(\eta_{n,m}) e^{-\frac{1}{4}|\eta_{n,m}|^2}=0.
\end{equation}
Since $G(\eta_{n,m})$ takes the value $-1$ on a lattice with twice the original
lattice constants, we may rewrite this as
\begin{equation}
  \label{eq:genfun1}
  \sum_{n,m} e^{\frac{1}{2}\eta_{n,m} \bar z} e^{-\frac{1}{4}|\eta_{n,m}|^2}-
  2\sum_{n,m} e^{\eta_{n,m} \bar z} e^{-|\eta_{n,m}|^2}=0.
\end{equation}
Kalmeyer and Laughlin~\cite{kalmeyer-89prb11879} observed that for the square
lattice, the second sum in \eqref{eq:genfun1} can be expressed as a sum of the
Fourier transform of the function we sum over in the first term.  We
demonstrate here that their proof can be extended to arbitrary lattices.

To begin with, we define the Fourier transform in complex coordinates
\begin{equation}
  \label{eq:ft}
  \tilde f(\zeta)=\int d^2\eta f(\eta) e^{i\Re(\eta\bar\zeta)},
\end{equation}
where $\Re$ denotes the real part and we have used \eqref{eq:area}.  Since the
area of the unit cell of our lattice is taken to be $2\pi$, the reciprocal
lattice is given by the original lattice rotated by $\frac{\pi}{2}$ in the
plane without any rescaling of the lattice constants.  In complex coordinates,
\begin{equation}
  \label{eq:reciprocal}
  \zeta_{n',m'}=i(n'a+m'b),
\end{equation}
as this immediately implies
\begin{eqnarray}
  \label{eq:reciprocalcheck}
  \bs{R}_{n,m}\cdot \bs{K}_{n',m'}\nonumber
  &=&\Re(\eta_{n,m}\bar\zeta_{n',m'})=\\\nonumber
  &=&\Re\left((na+mb)(-i)(n'\bar a+m'\bar b)\right)\\\nonumber
  &=&n m'\Im(a\bar b)+m n'\Im(b\bar a)\\\nonumber
  &=&2\pi\cdot\text{integer}.
\end{eqnarray}
Then 
\begin{equation}
  \label{eq:sumsum}
 \sum_{n',m'}\tilde f(\zeta_{n',m'}) = \Omega\sum_{n,m} f(\eta_{n,m}).
\end{equation}
Eq.\ \eqref{eq:sumsum} follows directly from
\begin{equation}
  \label{eq:delta}
  \sum_{n',m'} e^{i\Re(\eta\bar\zeta_{n',m'})}
    = \Omega\sum_{n,m} \delta^{(2)}(\eta_{n,m}-\eta),
\end{equation}
which is just the 2D equivalent of the (Dirac comb) identity
\begin{equation}
  \label{eq:delta1d}
  \sum_{n'=-\infty}^\infty e^{2\pi i n' x}
    = \sum_{n=-\infty}^\infty  \delta (x-n)
\end{equation}
The r.h.s.\ of \eqref{eq:delta1d} is obviously zero if $x$ is not an
integer, and manifestly periodic in x with period 1.  To verify
the normalization, observe that since for any $N$ odd,
\begin{equation}
  \label{eq:simplesum}\nonumber
  \sum_{n'=-\frac{N-1}{2}}^{+\frac{N-1}{2}}e^{2\pi i n' y/N}
  =\begin{cases}
    N &\text{for}\ y=N\cdot\text{integer}\\[4pt]
    \,0 &\text{otherwise}.
  \end{cases}
\end{equation}
This implies
\begin{equation}
  \label{eq:simplesum2}\nonumber
  \frac{1}{N}\sum_{y=-\frac{N-1}{2}}^{+\frac{N-1}{2}}
  \sum_{n'=-\frac{N-1}{2}}^{+\frac{N-1}{2}}e^{2\pi i n' y/N}=1,
\end{equation}
which in the limit $N\rightarrow\infty$ is equivalent to
\begin{equation}
  \label{eq:simplesum3}\nonumber
  \int_{-\frac{N}{2}}^{+\frac{N}{2}}\frac{dy}{N}
  \sum_{n'=-\frac{N-1}{2}}^{+\frac{N-1}{2}}e^{2\pi i n' y/N}=1
\end{equation}
Substituting $x=y/N$ yields
\begin{equation}
  \label{eq:simplesum4}\nonumber
  \int_{-\frac{1}{2}}^{+\frac{1}{2}}dx
  \sum_{n'=-\infty}^\infty e^{2\pi i n' x}=1,
\end{equation}
which proves the normalization in \eqref{eq:delta1d}. 

We proceed by evaluation of the Fourier transform of 
$f(\eta)=e^{\frac{1}{2}\eta \bar z} e^{-\frac{1}{4}|\eta|^2}$:
\begin{eqnarray}
  \label{eq:FTofeta}\nonumber
  \tilde f(\zeta)
  &=&\int d^2\eta\, e^{\frac{1}{2}\eta \bar z} e^{-\frac{1}{4}|\eta|^2} 
  e^{i\Re(\eta\bar\zeta)}\\\nonumber
  &=&\int d^2\eta\, e^{\frac{1}{2}\eta \bar z} e^{-\frac{1}{4}|\eta|^2} 
  e^{\frac{i}{2}(\eta\bar\zeta+\bar\eta\zeta)}\\ 
  &=&4\pi e^{-|\zeta|^2 +i\zeta\bar z}
\end{eqnarray}
where we have used the integral
\begin{eqnarray}
  \label{eq:gaussintegral}\nonumber
  \int d^2\eta\!&F(\eta)\!&  e^{-\frac{1}{\alpha}(|\eta|^2- \bar\eta w)}
  \\\nonumber
  &=&\!F(\alpha \partial_{\bar w})
  \int d^2\eta\, e^{-\frac{1}{\alpha}(|\eta|^2- \bar\eta w-\eta\bar w)}
  \biggl|_{\bar w=0}\biggr.  \\\nonumber
  &=&\!F(\alpha \partial_{\bar w})
  \int d^2\eta\, e^{-\frac{1}{\alpha}(|\eta-w|^2-w\bar w)} 
  \biggl|_{\bar w=0}\biggr.\\\nonumber
  &=&\!\alpha\pi\, F(\alpha \partial_{\bar w})
  e^{\frac{1}{\alpha}w\bar w}\;=\;\alpha\pi\, F(w) \biggl|_{\bar w=0}\biggr.
\end{eqnarray}
with $F(\eta)=e^{\frac{1}{2}\eta\bar z + \frac{i}{2}\eta\bar\zeta}$,
$\alpha=4$, and $w=2i\zeta$.

Substituting \eqref{eq:FTofeta} into \eqref{eq:sumsum} we obtain
\begin{eqnarray}
  \label{eq:substituted}
  \sum_{n,m} f(\eta_{n,m}) = 2\sum_{n',m'}  e^{-|\zeta_{n',m'}|^2 +i\zeta_{n',m'}\bar z}
\end{eqnarray}
If we now substitute $n'= -n$, $m'= -m$, and hence
$i\zeta_{n',m'}=\eta_{n,m}$ into the r.h.s.\ of \eqref{eq:substituted},
we obtain \eqref{eq:genfun1}.  This completes the proof.


\vspace{10pt} 

\emph{Generalization to lattices where the origin does not coincide with a
  lattice site.}---We now assume a shifted lattice with the sites given by
\begin{equation}
  \label{eq:sites}
  \eta_{n,m}=na+mb+c,
\end{equation}
where $n$ and $m$ are integer and $a$, $b$, and $c$ are complex numbers such
that the area of the unit cell $\Omega$ spanned by the primitive lattice
vectors $a$ and $b$ remains set to $2\pi$ (see \eqref{eq:area} above).
Then for any polynomial $P$ of $\eta_{n,m}$,
\begin{equation}
  \label{eq:perelc}
  \sum_{n,m} P(\eta_{n,m}) G(\eta_{n,m}) e^{-\frac{1}{4}|\eta_{n,m}|^2}=0,
\end{equation}
where the gauge factor is now given by
\begin{equation}
  \label{eq:gperelc}
  G(\eta_{n,m})= (-1)^{(n+1)(m+1)} e^{-\frac{\i}{2}\Im(\eta_{n,m} \bar c)}. 
\end{equation}

\vspace{3pt} \emph{Proof.}---With $\eta_{n,m}'=\eta_{n,m}-c$, we write
the exponential in \eqref{eq:perelc} as
\begin{align}
  e^{-\frac{1}{4}|\eta_{n,m}|^2} 
  &=e^{-\frac{1}{4}|\eta_{n,m}'|^2}
  e^{-\frac{1}{4}(\eta_{n,m}'\bar c + \bar\eta_{n,m}'c)}\, e^{-\frac{1}{4}|c|^2}
  \nonumber\\[0.2\baselineskip]
  &=e^{-\frac{1}{4}|\eta_{n,m}'|^2} e^{\frac{1}{4}(\eta_{n,m}'\bar c - \bar\eta_{n,m}'c)}\, 
  e^{-\frac{1}{2}\eta_{n,m}'\bar c}\, e^{-\frac{1}{4}|c|^2}
  \nonumber\\[0.2\baselineskip]
  &=e^{-\frac{1}{4}|\eta_{n,m}'|^2} e^{\frac{\i}{2}\Im(\eta_{n,m} \bar c)}\, 
  e^{-\frac{1}{2}\eta_{n,m}'\bar c}\, e^{-\frac{1}{4}|c|^2}\nonumber
\end{align}
If we now absorb the last two factors into the Polynomial $P(\eta_{n,m})$,
\eqref{eq:perelc} with \eqref{eq:gperelc} reduces to \eqref{eq:perel}.

\vspace{3pt} 
\emph{Comment.}---For most applications, it is convenient to write
\eqref{eq:gperelc} as
\begin{align}
  \label{eq:ggperelc}
  G(\eta_{n,m}) = (-1)^{(n+1)(m+1)}\,e^{\frac{\i}{2}[(a' n+b'm)c''-(a''n+b''m)c')]},
\end{align}
where $a=a'+\i a''$, $b=b'+\i b''$, $c=c'+\i c''$ and $a'$, $a''$,
etc.\ are real. 

\vspace{10pt} 
\emph{The Perelomov identity in uniaxial gauge.}---The generalized 
identity \eqref{eq:perelc} with \eqref{eq:gperelc} can further be rewritten as
\begin{equation}
  \label{eq:perelu}
  \sum_{n,m} P(\eta_{n,m}) G(\eta_{n,m})  e^{-\frac{1}{2}\Im(\eta_{n,m})^2}=0,
\end{equation}
with the gauge factor now given by 
\begin{equation}
  \label{eq:gperelu}
  G(\eta_{n,m})= (-1)^{(n+1)(m+1)} e^{\frac{\i}{2}\Re(\eta_{n,m}-c)\Im(\eta_{n,m}+c)}.
\end{equation}

\vspace{3pt} \emph{Proof.}---If we substitute  
\begin{align}
  P(\eta_{n,m})\to P(\eta_{n,m})\,e^{+\frac{1}{4}\eta_{n,m}^2}\nonumber
\end{align}
into \eqref{eq:perelc}, we obtain for the product of all the exponential factors
\begin{align}
  \hspace{10pt}&\hspace{-10pt}
  e^{-\frac{\i}{2}\Im(\eta_{n,m} \bar c)}\, e^{+\frac{1}{4}(\eta_{n,m}^2-|\eta_{n,m}|^2)}
  \nonumber\\[0.2\baselineskip]
  &=e^{-\frac{\i}{2}\Im(\eta_{n,m} \bar c)}\,
  e^{+\frac{1}{4}\eta_{n,m}(\eta_{n,m}-\bar\eta_{n,m})}
  \nonumber\\[0.2\baselineskip]
  &=e^{-\frac{\i}{2}\Im(\eta_{n,m} \bar c)}\,
  e^{+\frac{\i}{2}\left[\Re(\eta_{n,m})+\i\Im(\eta_{n,m})\right]\Im(\eta_{n,m})}
  \nonumber\\[0.2\baselineskip]
  &=e^{-\frac{\i}{2}\Im(\eta_{n,m} \bar c)}\,
  e^{+\frac{\i}{4}\Im(\eta_{n,m}^2)}\, e^{-\frac{1}{2}\Im(\eta_{n,m})^2}
  \nonumber\\[0.2\baselineskip]
  &=e^{+\frac{\i}{4}\Im[\eta_{n,m}(\eta_{n,m}-2\bar c)]}\, e^{-\frac{1}{2}\Im(\eta_{n,m})^2}
  \nonumber\\[0.2\baselineskip]
  &= e^{-\frac{\i}{4}\Im(\bar c^2)}\,
  e^{+\frac{\i}{4}\Im[(\eta_{n,m}-\bar c)^2]}\,e^{-\frac{1}{2}\Im(\eta_{n,m})^2}
  \nonumber\\[0.2\baselineskip]
  &= e^{+\frac{\i}{4}\Im(c^2)}\,
  e^{\frac{\i}{2}\Re(\eta_{n,m}-c)\Im(\eta_{n,m}+c)}\,
  e^{-\frac{1}{2}\Im(\eta_{n,m})^2}.
  \nonumber
\end{align}
If we absorb the first factor into the polynomial, we obtain
\eqref{eq:perelu} with \eqref{eq:gperelu}.

\vspace{3pt} \emph{Comment.}---For most applications, it is convenient
to write \eqref{eq:gperelu} as
\begin{align}
  \label{eq:ggperelu}
  G(\eta_{n,m}) = (-1)^{(n+1)(m+1)}\,e^{\frac{\i}{2}(a' n+b'm)(a''n+b''m+2c'')}.
\end{align}
For a rectangular lattice with $a''=b'=c''=0$, 
this gauge factor reduces to 
\begin{align}
  G(\eta_{n,m}) = (-1)^{n+m+1}.\\\nonumber
\end{align}

\vspace{-3pt}
\emph{Acknowledgments.}---We wish to thank W.~Lang for his critical
reading of parts of this manuscript.  MG is supported by the German
Research Foundation under grant FOR 960.  RT is supported by an SITP
fellowship at Stanford University.


\begin{thebibliography}{10}

\bibitem{anderson87s1196}
P.~W. Anderson, Science {\bf 235},  1196  (1987).

\bibitem{kalmeyer-87prl2095}
V. Kalmeyer and R.~B. Laughlin, Phys. Rev. Lett. {\bf 59},  2095  (1987).

\bibitem{kivelson-87prb8865}
S.~A. Kivelson, D.~S. Rokhsar, and J.~P. Sethna, Phys. Rev. B {\bf 35},  8865
  (1987).

\bibitem{wen-89prb11413}
X.~G. Wen, F. Wilczek, and A. Zee, Phys. Rev. B {\bf 39},  11413  (1989).

\bibitem{zou-89prb11424}
Z. Zou, B. Doucot, and B.~S. Shastry, Phys. Rev. B {\bf 39},  11424  (1989).

\bibitem{kalmeyer-89prb11879}
V. Kalmeyer and R.~B. Laughlin, Phys. Rev. B {\bf 39},  11879  (1989).

\bibitem{moessner-01prl1881}
R. Moessner and S.~L. Sondhi, Phys. Rev. Lett. {\bf 86},  1881  (2001).

\bibitem{balents-02prb224412}
L. Balents, M.~P.~A. Fisher, and S.~M. Girvin, Phys. Rev. B {\bf 65},  224412
  (2002).

\bibitem{greiter02jltp1029}
M. Greiter, J.~Low Temp. Phys. {\bf 126},  1029  (2002).

\bibitem{kitaev06ap2}
A. Kitaev, Ann.~of Phys. {\bf 321},  2  (2006).

\bibitem{schroeter-07prl097202}
D.~F. Schroeter, E. Kapit, R. Thomale, and M. Greiter, Phys. Rev. Lett. {\bf
  99},  097202  (2007).

\bibitem{yao-07prl247203}
H. Yao and S.~A. Kivelson, Phys. Rev. Lett. {\bf 99},  247203  (2007).

\bibitem{dusuel-08prb125102}
S. Dusuel, K.~P. Schmidt, J. Vidal, and R.~L. Zaffino, Phys. Rev. B {\bf 78},
  125102  (2008).

\bibitem{lee08s1306}
P.~A. Lee, Science {\bf 321},  1306  (2008).

\bibitem{greiter-09prl207203}
M. Greiter and R. Thomale, Phys. Rev. Lett. {\bf 102},  207203  (2009).

\bibitem{thomale-09prb104406}
R. Thomale, E. Kapit, D.~F. Schroeter, and M. Greiter, Phys. Rev. B {\bf 80},
  104406  (2009).

\bibitem{hermele-09prl135301}
M. Hermele, V. Gurarie, and A.~M. Rey, Phys. Rev. Lett. {\bf 103},  135301
  (2009).

\bibitem{balents10n199}
L. Balents, Nature {\bf 464},  199  (2010).

\bibitem{zhang-11prb075128}
Y. Zhang, T. Grover, and A. Vishwanath, Phys. Rev. B {\bf 84},  075128  (2011).

\bibitem{grover-11prl077203}
T. Grover and T. Senthil, Phys. Rev. Lett. {\bf 107},  077203  (2011).

\bibitem{yao-11prl087205}
H. Yao and D.-H. Lee, Phys. Rev. Lett. {\bf 107},  087205  (2011).

\bibitem{scharfenberger-11prb140404}
B. Scharfenberger, R. Thomale, and M. Greiter, Phys. Rev. B {\bf 84},  140404
  (2011).

\bibitem{nielsen-1201.3096}
A.~E.~B. Nielsen, J.~I. Cirac, and G. Sierra, arXiv:1201.3096.

\bibitem{greiter-1201.5312}
M. Greiter, D.~F. Schroeter, and R. Thomale, arXiv:1201.5312.

\bibitem{meng-10n847}
Z.~Y. Meng, T.~C. Lang, S. Wessel, F.~F. Assaad, and A. Muramatsu, Nature {\bf
  464},  847  (2010).

\bibitem{jiang-11arXiv:1112.2241}
H.-C. Jiang, H. Yao, and L. Balents, arXiv:1112.2241.

\bibitem{wen89prb7387}
X.~G. Wen, Phys. Rev. B {\bf 40},  7387  (1989).

\bibitem{wen90ijmp239}
X.~G. Wen, Int. J. Mod. Phys. {\bf B4},  239  (1990).

\bibitem{wen-90prb9377}
X.~G. Wen and Q. Niu, Phys. Rev. B {\bf 41},  9377  (1990).

\bibitem{wen04}
X. Wen, {\em Quantum Field Theory of Many-Body Systems}, {\em Oxford Graduate
  Texts} (Oxford University, New York, 2004).

\bibitem{levin-05prb045110}
M.~A. Levin and X.~G. Wen, Phys. Rev. B {\bf 71},  045110  (2005).

\bibitem{isakov-11np772}
S.~V. Isakov, M.~B. Hastings, and R.~G. Melko, Nature Physics {\bf 7},
  772–775  (2011).

\bibitem{moore10n194}
J.~E. Moore, Nature {\bf 464},  194  (2010).

\bibitem{hasan-10rmp3045}
M.~Z. Hasan and C.~L. Kane, Rev. Mod. Phys. {\bf 82},  3045  (2010).

\bibitem{qi-11rmp1057}
X.-L. Qi and S.-C. Zhang, Rev. Mod. Phys. {\bf 83},  1057  (2011).

\bibitem{laughlin-90prb664}
R.~B. Laughlin and Z. Zou, Phys. Rev. B {\bf 41},  664  (1990).

\bibitem{laughlinPC}
R.~B. Laughlin, private communication.

\bibitem{perelomov71tmp156}
A.~M. Perelomov, Theoret. Math. Phys. {\bf 6},  156  (1971).

\bibitem{moore-91npb362}
G. Moore and N. Read, Nucl. Phys. B {\bf 360},  362  (1991).

\bibitem{greiter-92npb567}
M. Greiter, X.~G. Wen, and F. Wilczek, Nucl. Phys. B {\bf 374},  567  (1992).

\bibitem{read-99prb8084}
N. Read and E. Rezayi, Phys. Rev. B {\bf 59},  8084  (1999).

\end{thebibliography}

\end{document}